\documentclass[usenatbib]{mn2e}
\usepackage{color,graphics,graphicx,float,fixltx2e,natbib,epsfig,mathptmx,relsize}
\usepackage{savesym}
\usepackage{amsmath}
\savesymbol{iint}
\savesymbol{iiint}
\usepackage{txfonts}
\restoresymbol{TXF}{iint}
\restoresymbol{TXF}{iiint}

\makeatletter
\def\@cite#1#2{\textsuperscript{[{#1\if@tempswa , #2\fi}]}}
\makeatother
\makeatletter
\newcommand*{\rom}[1]{\expandafter\@slowromancap\romannumeral #1@}
\makeatother

\def\ion#1#2{#1$\;${\smaller\rm\rom{#2}}\relax}

\begin{document}

\title[XMM-Newton Large Program on SN1006 - II: Thermal Emission]{\emph{XMM-Newton} Large Program on SN1006 - II: Thermal Emission}
\author[J. T. Li et al.]{Jiang-Tao Li$^{1}$\thanks{E-mail:pandataotao@gmail.com}, Anne Decourchelle$^{2}$, Marco Miceli$^{3,5}$, Jacco Vink$^{4}$, and Fabrizio Bocchino$^{5}$\\
$^{1}$Department of Astronomy, University of Michigan, 311 West Hall, 1085 S. University Ave, Ann Arbor, MI, 48109-1107, U.S.A.\\
$^{2}$Service d'Astrophysique, CEA Saclay, 91191 Gif-sur-Yvette Cedex, France\\
$^{3}$Dipartimento di Fisica \& Chimica, Universit$\rm\grave{a}$ di Palermo, Piazza del Parlamento 1, I-90134 Palermo, Italy\\
$^{4}$Anton Pannekoek Institute/GRAPPA, University of Amsterdam, PO Box 94249, 1090 GE Amsterdam, The Netherlands\\
$^{5}$INAF-Osservatorio Astronomico di Palermo, Piazza del Parlamento, 90134 Palermo, Italy}

\maketitle

\begin{abstract}
Based on the \emph{XMM-Newton} large program on SN1006 and our newly developed tools for spatially resolved spectroscopy analysis as described in \citet{Li15} (Paper~I), we study the thermal emission from ISM and ejecta of SN1006 by analyzing the spectra extracted from 583 tessellated regions dominated by thermal emission. With some key improvements in spectral analysis as compared to Paper~I, we obtain much better spectral fitting results with significantly less residuals. The spatial distributions of the thermal and ionization states of the ISM and ejecta show significantly different features, which are in general consistent with a scenario that the ISM (ejecta) is heated and ionized by the forward (reverse) shock propagating outward (inward). Different heavy elements show different spatial distributions so different origins, with Ne mostly from the ISM, Si and S mostly from the ejecta, and O and Mg from both the ISM and ejecta. Fe L-shell line emissions are only detected in a small shell-like region SE to the center of SN1006, indicating that most of the Fe-rich ejecta has not yet or just recently been reached by the reverse shock. The overall abundance patterns of the ejecta for most of the heavy elements, except for Fe and sometimes for S, are consistent with typical Type~Ia SN products. The NW half of the SNR interior between the NW shell and the soft X-ray brighter SE half probably represents a region with turbulently mixed ISM and ejecta, so has enhanced emission from O, Mg, Si, and S, lower ejecta temperature, and a large diversity of ionization age. In addition to the asymmetric ISM distribution, an asymmetric explosion of the progenitor star is also needed to explain the asymmetric ejecta distribution. 
\end{abstract}


\section{Introduction}\label{PaperIIsec:Introduction}

The thermal X-ray emission from young supernova (SN) remnants (SNRs) is often comprised of different components. These X-ray emission components are produced by the interstellar medium (ISM) and SN ejecta shocked by the SNR blast wave or the reverse shock. These shocked plasma components mix with each other either physically or in projection, producing different thermal, chemical, and ionizational states in the observed X-ray spectra. Decomposing these plasma components plays an important role in understanding the under-ionized plasma, energy non-equipartition between different particles, and shock history of the post-shock gas.

The historical Type~Ia SNR SN1006 (e.g., \citealt{Stephenson10}) is one of the best cases suitable for the study of the spatial distributions of different hot plasma components. \emph{First}, owing to its high Galactic latitude ($b=14.6^\circ$), the foreground extinction to SN1006 is relatively low ($N_{\rm H}=6.8\times10^{20}\rm~cm^{-2}$; \citealt{Dubner02}) among the Galactic SNRs. Therefore, soft X-ray emission lines from some relatively light elements, for example, oxygen, are particularly strong compared to other young SNRs of the same type (e.g., the Tycho's SNR; \citealt{Decourchelle01}; G1.9+0.3; \citealt{Borkowski13}).  \emph{Second}, the density of the surrounding medium of SN1006 is low ($n_{\rm 0}\sim0.3\rm~cm^{-3}$; \citealt{Dubner02}). As a result, SN1006 likely stays in the earliest evolutionary stage among the few historical SNRs (Tycho, Kepler, Cas~A, etc.), although it is already over a thousand years old. This is another key reason of the particularly strong oxygen lines, because most of the oxygen atoms have not yet been completely ionized. The shocked plasma properties of SN1006 show significant dispersion, and some heavy elements, such as Fe, even appear to be newly shocked (e.g., \citealt{Yamaguchi08}). SN1006 is thus a potential candidate for studying the shock heating and ionization of the ISM and the SN ejecta at very early stage of a SNR's hydrodynamic evolution. \emph{Third}, the distance to SN1006 is just $\approx2.18\rm~kpc$ \citep{Winkler03}. As a result, the X-ray image shows many well resolved features (e.g., \citealt{Miceli09,Uchida13,Winkler14}) which are large and bright enough for spatially resolved spectral analysis.

In \citet{Li15} (Paper~I), we have conducted spatially resolved spectroscopy analysis of SN1006 based on our newly developed tools, as well as our \emph{XMM-Newton} Large Program (LP) and archival data. We have constructed images of many parameters based on spectral modeling of 3596 tessellated regions with a 1-T model plus a non-thermal synchrotron component and various background components. In particular, we found that the Fe abundance estimated based on the 1-T model is enhanced only in a small region located off the geometric center and to the southeast (SE) of the SNR. This strongly suggests that the Fe-rich ejecta is newly shocked, consistent with the low ionization state of Fe lines as revealed in both X-ray emission lines (e.g., \citealt{Yamaguchi08,Uchida13}) and optical/UV absorption lines from background sources (e.g., \citealt{Wu93,Hamilton97,Winkler05}). Furthermore, we also found multiple peaks in the probability distribution functions (PDFs) of the temperature ($kT$) and ionization parameter ($n_{\rm e}t$) of the single-temperature plasma. This indicates that even in regions as small as a few tens of arcseconds (the typical size of the tessellated regions as adopted in Paper~I), there could be significant variations of the thermal and ionizational states of the plasma. We also found some significant residuals in the spectral fitting of some regions. This indicates that further improvement in spectral decomposition and modeling are needed.

In the present paper, we will decompose the thermal emission of SN1006 into the ISM and ejecta components and further study their spatial distributions. The paper is organized as follows: in \S\ref{PaperIIsec:ThermalDataAnalysis}, we describe several improvements of the spectral analysis procedures as adopted in Paper~I and the new ``2-T'' model used to decompose the ISM and ejecta components. In \S\ref{PaperIIsec:Discussion}, we present the key results based on the updated analysis of the thermal emission of SN1006 and discuss their scientific implications. Our main results and conclusions are summarized in \S\ref{PaperIIsec:Summary}.

\section{Spatially resolved spectral analysis of the regions dominated by thermal emission}\label{PaperIIsec:ThermalDataAnalysis}

\subsection{Tessellated meshes and spectra extraction}\label{PaperIIsubsec:MeshSpectra}

The \emph{XMM-Newton} large program on SN1006, the processes of data reduction, and the methods of spatially resolved spectroscopy analysis have been described in detail in Paper~I.

\begin{figure}
\begin{center}
\hspace{-0.2in} 
\epsfig{figure=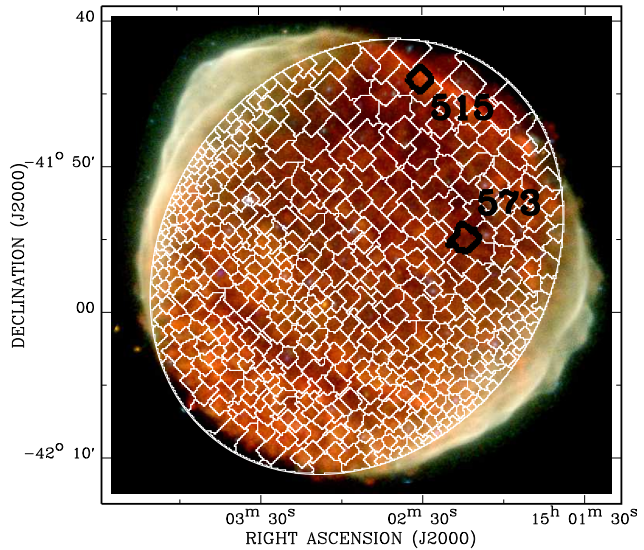,width=0.5\textwidth,angle=0, clip=}
%\vspace{-0.3in} 
\caption{583 tessellated meshes used for spatially resolved spectroscopy analysis overlaid on the tricolor images of SN1006 (red: 0.3-1~keV; green: 1-2~keV; blue: 2-8~keV). Each region contains $\gtrsim 10^3\rm~counts$ in 1.66-1.95~keV from the combination of MOS-1, MOS-2, and PN. The two regions marked in black are used to extract the example spectra shown in Fig.~\ref{PaperIIfig:samplespec}.}\label{PaperIIfig:meshes}
\end{center}
\end{figure}

In Paper~I, we conducted spatially resolved spectral analysis in 3596 tessellated regions each contains $\gtrsim 10^4\rm~counts$ in 0.3-8~keV. This high resolution run is sufficient to decompose the thermal and non-thermal components in the spectra and to roughly characterize the average thermal (traced by the electron temperature $kT$) and ionization states (traced by the ionization parameter $n_{\rm e}t$) of the plasma. However, the tessellated regions in Paper~I typically have too few photons to resolve the Si-K$\alpha$ lines at $\sim1.8\rm~keV$, which are known to be broader than expected from a single non-equilibrium ionization (NEI) model in SN1006 (e.g., \citealt{Yamaguchi08,Uchida13}), therefore plays a key role in studying the diversity of the thermal and ionization states of the plasma. Furthermore, in order to save computer time, we have stacked the spectra extracted from different observations in Paper~I. Because there is a deficiency in the low-energy calibration of the PN camera, the energy response of the low energy spectra of PN is changing with time \citep{Dennerl04}. As a result, the direct stacking of the spectra taken at different times causes significant residual in fitting the low energy spectra of PN (typically at $\lesssim1\rm~keV$). This is the main reason of the large $\chi^2/\rm d.o.f.$ (often $>2.0$) of the spectra dominated by thermal emission in Paper~I (the residuals are mainly in the energy range dominated by the \ion{O}{7} and \ion{O}{8} lines).

\begin{figure*}
\begin{center}
\hspace{-0.25in} 
\epsfig{figure=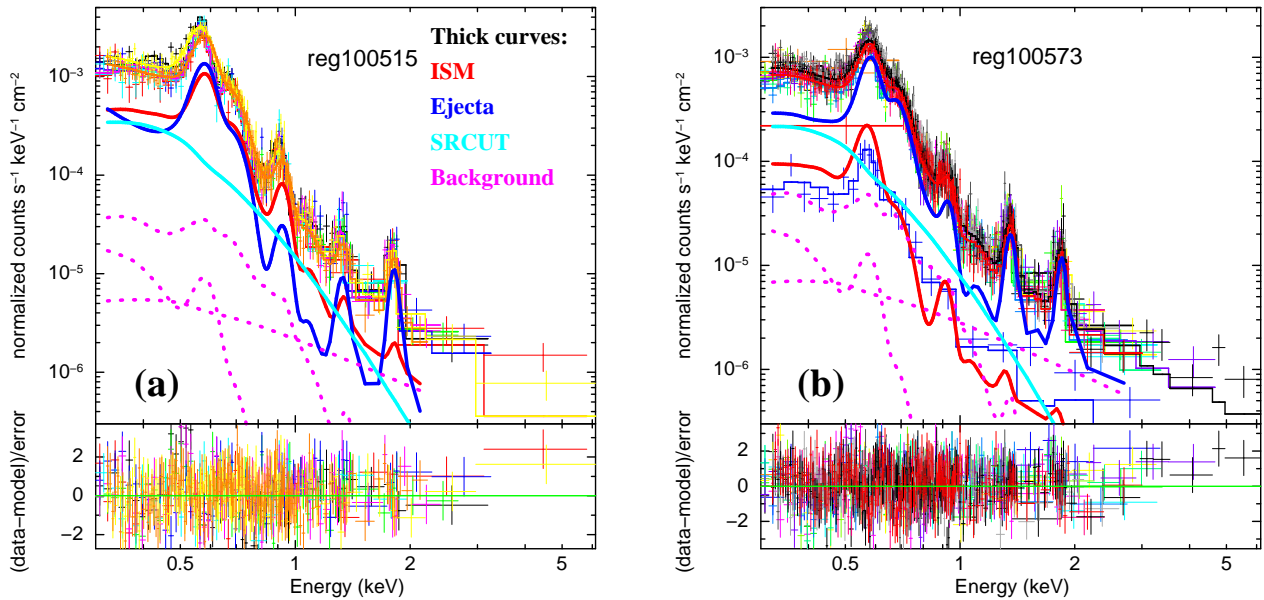,width=1.0\textwidth,angle=0,clip=}
\caption{Example spectra extracted from the two regions marked in black in Fig.~\ref{PaperIIfig:meshes}. Panel~(a) shows an example with comparable contributions from the ISM and ejecta components, while Panel~(b) shows an example dominated by the ejecta emission. The colored data points are the instrumental background-subtracted spectra extracted from different instruments (MOS-1, MOS-2, and PN) and different observations, while the thin solid curves with the corresponding colors are their best-fit models (often not visible). Different model components are plotted as thick solid curves, as denoted on the top right corner of Panel~(a). The Gaussian lines have too low flux to show up on both panels. The magenta dotted curves are the sky background components as detailed in the appendix of Paper~I. The lower panels show the residuals of the fitting. The $\chi^2/\rm d.o.f.$ for the two panels are 1.168 (704.14/603) and 1.127 (1417.55/1258), respectively. Some of the best fit parameters and the corresponding errors are summarized in Table~\ref{table:paraexample}.}\label{PaperIIfig:samplespec}
\end{center}
\end{figure*}

In the present paper, we improve the spectral analysis procedures as adopted in Paper~I in the following ways. \emph{Firstly}, we focus on the thermal emission from the interior region of SN1006. We filter out the northeast (NE) and southwest (SW) limbs dominated by non-thermal emission using an elliptical region shown in Fig.~\ref{PaperIIfig:meshes}. It is generally difficult to study the properties of different thermal emission components in the two limbs dominated by non-thermal emission. \emph{Secondly}, we re-construct the tessellated meshes to extract spectra with higher signal-to-noise ratios. We set a threshold that each mesh constains $\gtrsim 10^3\rm~counts$ in the energy range of the Si~K bump (1.66-1.95~keV, typically including emission lines from \ion{Si}{9} to \ion{Si}{13}; Paper~I). The spectra extracted from each tessellated regions are thus typically sufficient to resolve all the prominent soft X-ray emission lines (\ion{O}{7}, \ion{O}{8}, Ne, Mg, Si, and in many cases S). 583 regions are created within the elliptical region dominated by thermal emission (Fig.~\ref{PaperIIfig:meshes}). We call the analysis based on these meshes the ``low-resolution run'', in contrast to the ``high-resolution run'' presented in Paper~I. \emph{Thirdly}, we create the Redistribution Matrix File (RMF) for all the spectra extracted from the three instruments (MOS-1, MOS-2, and PN) of the 15 observations and the 583 regions, instead of using the templet response files in Paper~I. \emph{Finally}, we jointly fit the spectra extracted from different observations instead of stacking them and directly fitting the stacked spectra. Examples of the spectra extracted from individual meshes are shown in Fig.~\ref{PaperIIfig:samplespec}, which show significant \ion{O}{7}, \ion{O}{8}, Ne, Mg, Si emission line features and some weak residuals at higher energy probably from emission lines of heavier elements.

\subsection{Spectral model}\label{PaperIIsubsec:SpecModel}

The plasma of SN1006 is comprised of various components with different thermal, chemical, and ionizational states (e.g.,  \citealt{Yamaguchi08,Miceli12,Uchida13}). In Paper~I, we have adopted a simple {\small VNEI+SRCUT} model plus three {\small Gaussian} lines representing the \ion{O}{7} high level transitions not included in the {\small VNEI} code (K$\delta$, K$\epsilon$, K$\zeta$), in addition to several background components. This model is insufficient to decompose the various plasma components and often produce significant residuals in some prominent emission lines (e.g., the Si-K$\alpha$ lines at $\sim1.8\rm~keV$). We herein adopt a ``2-T'' model (two {\small VNEI}) representing the various plasma components, plus an {\small SRCUT} model to describe the residual non-thermal emission, three linked {\small Gaussian} lines representing the \ion{O}{7} K$\delta-\zeta$ transitions, and the same background components as adopted in Paper~I. The two {\small VNEI} components are assumed to be from the shocked ISM and ejecta, respectively. Both of them (also the {\small SRCUT} and {\small Gaussian} lines) are subject to the foreground extinction of $N_{\rm H}=6.8\times10^{20}\rm~cm^{-2}$ \citep{Dubner02}.

We fix most of the parameters of the ISM component. \emph{Firstly}, we assume solar abundance for all the heavy elements (using the abundance table from \citealt{Anders89}), consistent with an ISM assumption. \emph{Secondly}, we noticed that the temperature of the ISM component cannot be well constrained. This is because the ISM mostly contributes to the oxygen and neon lines at $<1\rm~keV$, and most of the higher energy lines are dominated by the ejecta component (e.g., Fig.~\ref{PaperIIfig:samplespec}). The relatively narrow energy range and small number of available emission lines makes the plasma temperature poorly constrained in most of the regions. Instead of directly determining it from spectral fitting, we fix the ISM temperature at 1.5~keV, the value obtained from a careful analysis of the southeast (SE) edge of the SNR \citep{Miceli12}. We caution that this assumption may be incorrect for some regions known to have lower temperature shocked ISM [e.g., the northwest (NW) rim; Paper~I], but will typically not affect our determination of the ejecta properties because the ISM component usually has much lower abundances than the ejecta and does not contribute significantly to the high energy emission lines such as those from Si and S (mostly from the ejecta). With most of the parameters fixed, there are only two free parameters of the ISM component: the ionization parameter $n_{\rm e}t_{\rm ISM}$ and the normalization.

For the ejecta component, most of the parameters are set free, including the temperature ($kT_{\rm ejecta}$), the ionization parameter $n_{\rm e}t_{\rm ejecta}$, the redshift (converted to the line of sight velocity $v_{\rm ejecta}$ in the following sections), the normalization, and the abundances of O, Ne, Mg, Si, S, and Fe. Paper~I reveals significant spatial variation of the O lines. We therefore let the O abundance to be freely variable. In X-ray spectral analysis, there is a degeneracy between the O abundance and the normalizaiton of the {\small VNEI}. Therefore, the fitted absolute value of the O abundance (and also the abundances of other heavy elements) may have large uncertainties. Furthermore, the degree of microscopic mixing between shocked ISM and ejecta is poorly known, so the absolute abundances of heavy elements are also poorly constrained theoretically. We have tested models with the O abundance fixed at different values (e.g., similar as those adopted in \citealt{Yamaguchi08,Uchida13}), but found no significant differences in the resultant abundance ratios between other heavy elements and O. In the following sections, we only present the abundance ratio between different elements and oxygen for quantitative discussions. We also caution that the accuracy of the absolute O abundance affects the determination of the emission measure (EM) and thus the estimation of the electron number density ($n_{\rm e,ejecta}$). Different from Paper~I, the S abundance is set free, independent on the Si abundance, as there are often some S line features which could be roughly characterized based on the high counting statistic at even $\gtrsim2\rm~keV$ (e.g., Fig.~\ref{PaperIIfig:samplespec}). Since the Fe~K-shell line features cannot be resolved in the spectra of most regions (they are detected in the spectra of the whole SNR; Paper~I), the Fe abundance is constrained with the L-shell lines which form a bump at $\sim1\rm~keV$. The Ni abundance is linked to the value of Fe, while the C, N, Ar, and Ca abundances are fixed at solar values because the emission lines from them are too weak in our spectra. 

The {\small Gaussian} lines representing the \ion{O}{7} high level transitions are also subject to the same redshift and foreground extinction as the {\small VNEI} component of the ejecta. Similar as in Paper~I, we assume the instrumental broadening is the only source of broadening for these lines. The centroid energy of the three lines are also fixed at 0.714, 0.723, and 0.730~keV, respectively. We further link the normalization ratio of them by K$\epsilon$/K$\delta$=K$\zeta$/K$\epsilon$=0.5 \citep{Yamaguchi08}. There is thus only one free parameter of these {\small Gaussian} lines: the normalization of the \ion{O}{7} K$\delta$ line.

The non-thermal and background components are the same as adopted in Paper~I. In particular, the normalization of the {\small SRCUT} is fixed at the value converted from the 1.4~GHz flux-accurate image of \citet{Dyer09}, while the photon index $\alpha$ and cutoff frequency $\nu_{\rm cutoff}$ are set free. All the background parameters are fixed and scaled from the background spectra presented in the Appendix of Paper~I.

\begin{table}
\begin{center}
\caption{Parameters and errors of the two example regions shown in Fig.~\ref{PaperIIfig:meshes}. Errors are statistical only and are quoted at 90\% confidence level.} %\scriptsize
\begin{tabular}{lcccccccccccccc}
\hline
Parameter & reg100515 & reg100573 \\
\hline
$\log (n_{\rm e}t/{\rm cm^{-3}~s})_{\rm ISM}$ & $9.320_{-0.103}^{+0.034}$ & $8.948_{-0.076}^{+0.220}$ \\
$n_{\rm e,ISM}/{\rm cm^{-3}}$ & $0.76_{-0.05}^{+0.07}$ & $0.21_{-0.02}^{+0.03}$ \\
$kT_{\rm ejecta}/{\rm keV}$ & $1.29_{-0.26}^{+0.25}$ & $>9.2$ \\
$\log (n_{\rm e}t/{\rm cm^{-3}~s})_{\rm ejecta}$ & $9.065_{-0.089}^{+0.070}$ & $9.532_{-0.021}^{+0.008}$ \\
$n_{\rm e,ejecta}/{\rm cm^{-3}}$ & $0.26_{-0.02}^{+0.05}$ & $0.27_{-0.01}^{+0.02}$ \\
$Z_{\rm O,ejecta}/\rm solar$ & $<20$ & $2.39_{-0.33}^{+0.74}$ \\
$(Z_{\rm Ne}/Z_{\rm O})_{\rm ejecta}$ & $0.62_{-0.19}^{+0.21}$ & $0.28_{-0.02}^{+0.03}$ \\
$(Z_{\rm Mg}/Z_{\rm O})_{\rm ejecta}$ & $3.78_{-0.91}^{+0.89}$ & $1.43_{-0.13}^{+0.12}$ \\
$(Z_{\rm Si}/Z_{\rm O})_{\rm ejecta}$ & $13.8_{-2.8}^{+9.7}$ & $3.22\pm0.33$ \\
$(Z_{\rm S}/Z_{\rm O})_{\rm ejecta}$ & $30.1_{-11.6}^{+9.1}$ & $2.21_{-0.95}^{+1.51}$ \\
$(Z_{\rm Fe}/Z_{\rm O})_{\rm ejecta}$ & $<0.39$ & $<0.13$ \\
$v_{\rm ejecta}/\rm (km~s^{-1})$ & $1238_{-344}^{+264}$ & $2818_{-443}^{+75}$ \\
$\alpha$ & $0.11_{-0.01}^{+0.06}$ & $0.1 (<0.103)$ \\
$\nu_{\rm cutoff}/\rm Hz$ & $7.51_{-1.22}^{+0.94}\times10^{14}$ & $6.54_{-0.09}^{+0.78}\times10^{14}$ \\
$\chi^2/\rm d.o.f.$ & 704.14/603 & 1417.55/1258 \\
\hline
\end{tabular}\label{table:paraexample}
\end{center}
\end{table}

As a key improvement from Paper~I, we jointly fit the spectra extracted from different observtions. In order to account for the deficiency in the low-energy calibration of the PN camera, we convolve a {\small Gain} model to each PN spectra. Similar as in Paper~I, the slope of the {\small Gain} is fixed at 1 and the offset of each observation is set free. We also multiply a constant normalization factor to all of the spectra, in order to account for the possible difference in area scale and calibration bias of the spectra extracted from different instruments and different observations.

Examples of the fitted spectra from individual regions are presented in Fig.~\ref{PaperIIfig:samplespec}, and the best-fit parameters are summarized in Table~\ref{table:paraexample}. Since calculating errors of spectral parameters is very time consuming, we do not finish error calculation in the spectral analysis of all the regions. Instead, we show the typical values of two example regions in Table~\ref{table:paraexample}.

\begin{figure}
\begin{center}
\epsfig{figure=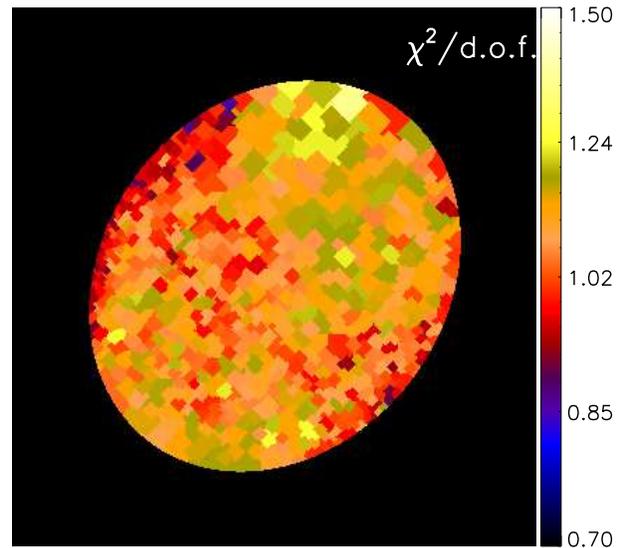,width=0.5\textwidth,angle=0, clip=}
\caption{$\chi^2/\rm d.o.f$ of the ``2-T'' joint spectral fit of the low-resolution run. The maximum $\chi^2/\rm d.o.f$ is 1.347, while the minimum value is 0.859.}\label{PaperIIfig:reducedchi}
\end{center}
\end{figure}

We have individually checked the spectra extracted from all the regions to confirm that the fitting is at least reasonable. The $\chi^2/\rm d.o.f.$ map is shown in Fig.~\ref{PaperIIfig:reducedchi}. Compared to the ``1-T'' fitting of the stacked spectra in Paper~I, the $\chi^2/\rm d.o.f.$ reduces significantly in this joint spectral analysis with a ``2-T'' model. The maximum $\chi^2/\rm d.o.f.$ in this ``2-T'' fit of the low-resolution run is 1.35, compared to 2.81 for the high resolution run with less spectral features in Paper~I.

Similar as in Paper~I, we also derive some parameters from the direct spectral fitting parameters. In particular, the electron number density $n_{\rm e}$ is derived from the normalization of the {\small VNEI}, using the same geometric model as Paper~I (originally from \citealt{Miceli12}). We caution that this shell-like geometric model is for the shocked ISM, which may not be optimized for the ejecta component. However, since the structure of the ejecta is not well constrained, we adopt the same geometric model as the ISM for a rough estimation of $n_{\rm e,ejecta}$. As mentioned above, another potential uncertainty in estimating $n_{\rm e}$ is the absolute abundance of oxygen, which shows a significant degeneracy with the normalization of the {\small VNEI}. For the ISM component, it is generally reliable to fix the oxygen abundance at solar, but for the ejecta, it makes a great difference whether the dominant species is hydrogen or oxygen (produces the most prominent emission features in soft X-ray). In this paper, we estimate $n_{\rm e,ejecta}$ based on the best-fit abundances of heavy elements. We further derive the ionization age $t_{\rm ion}\equiv n_{\rm e}t/n_{\rm e}$ from $n_{\rm e}$ and the directly fitted $n_{\rm e}t$. $t_{\rm ion}$ has truly time dimension and tracks the ionization history corrected for the local density variation.

\section{Results and Discussions}\label{PaperIIsec:Discussion}

\subsection{Spatial distributions of the ISM and ejecta components}\label{PaperIIsubsec:DiscussionSpatialDistribution}

\begin{figure*}
\begin{center}
\epsfig{figure=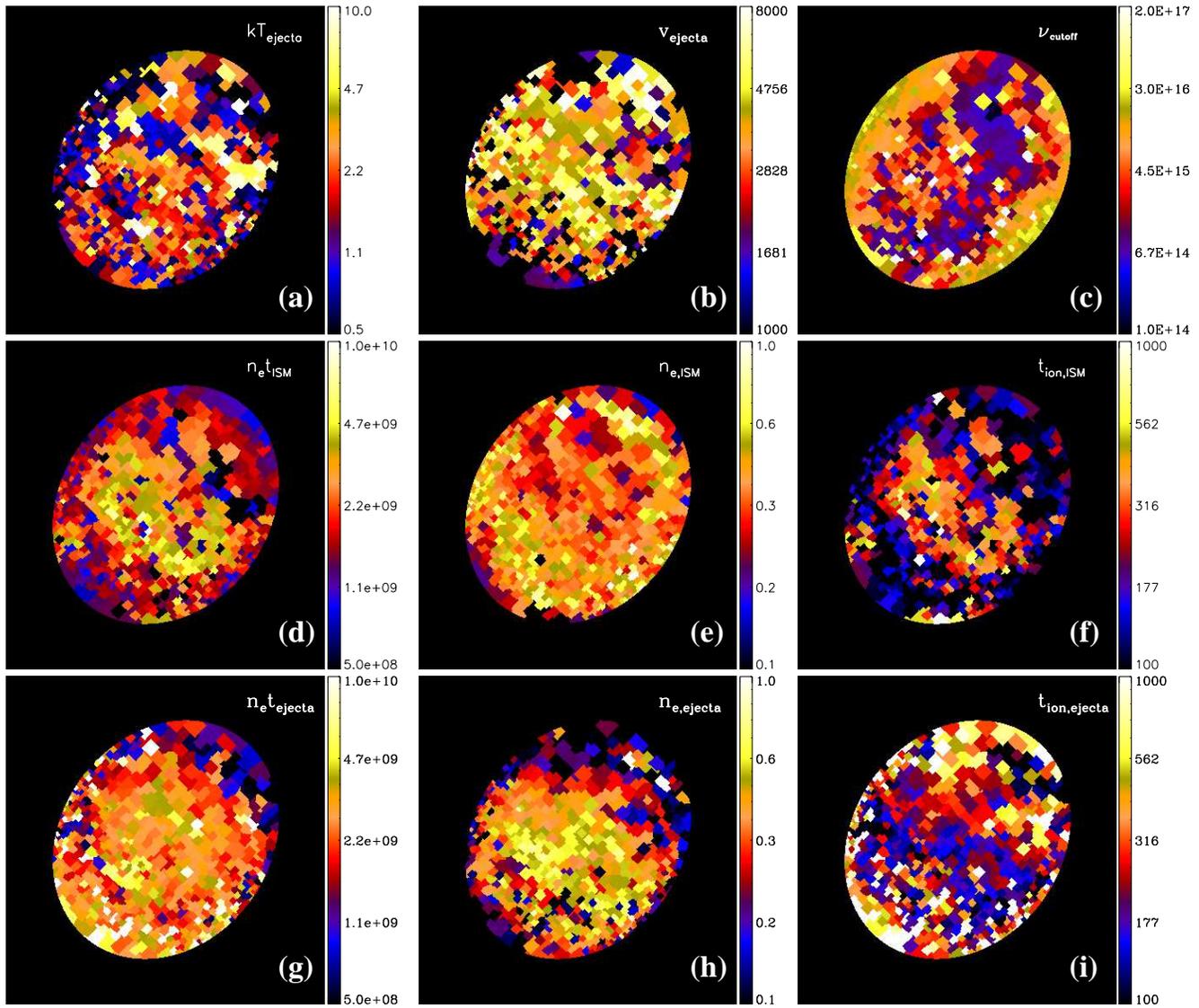,width=1.0\textwidth,angle=0, clip=}
\caption{Parameter maps constructed by jointly analyzing the spectra from the low-resolution meshes (Fig.~\ref{PaperIIfig:meshes}) with the ``2-T'' model. (a) Ejecta temperature in keV. (b) Line of sight velocity (positive toward the observer) in $\rm km~s^{-1}$. (c) Cutoff frequency of the non-thermal SRCUT component in Hz. (d-f) The ionization parameter in $\rm cm^{-3}~s$, electron number density in $\rm cm^{-3}$, and ionization age in year of the ISM component. (g-i) Similar as (d-f), but for the ejecta component.}\label{PaperIIfig:paraimg}
\end{center}
\end{figure*}

In Fig.~\ref{PaperIIfig:paraimg}, we present some of the parameter maps based on the joint spectral analysis with the ``2-T'' model of the low-resolution run.

In general, all the parameters are in reasonable ranges, indicating that the systematical analysis of the spectra is reliable. In particular, the temperature of the ejecta (Fig.~\ref{PaperIIfig:paraimg}a) is typically higher ($\sim2.5\rm~keV$) than obtained from the high-resolution run in Paper~I (typically $\sim1.5\rm~keV$ in the same region). This is because the Si lines are now included in the spectral fitting, which often need a high temperature component (in most of the cases the ejecta) to describe. The line of sight velocity of the ejecta ($v_{\rm ejecta}$; blue shifted defined as positive; Fig.~\ref{PaperIIfig:paraimg}b) has the highest value in the center and decreases from inner to outer regions, consistent with an expanding shell geometry with most of the soft X-ray emission contributed by the ejecta in the near side. The typical value of $v_{\rm ejecta}$ is $<5000\rm~km~s^{-1}$, also consistent with the measured proper motion of the non-thermal filaments \citep{Winkler14}. The contribution of non-thermal emission, as indicated by the cutoff frequency of the {\small SRCUT} component (Fig.~\ref{PaperIIfig:paraimg}c; typically $<3\times10^{16}\rm~Hz$), is much smaller than the NE and SW limbs ($\nu_{\rm cutoff}$ is typically $>10^{17}\rm~Hz$ on the non-thermal filaments; Paper~I). 

The ionization age and electron number density ($n_{\rm e}t$, $n_{\rm e}$, and $t_{\rm ion}$) of the ISM (Fig.~\ref{PaperIIfig:paraimg}d-f) and ejecta (Fig.~\ref{PaperIIfig:paraimg}g-i) are roughly in the same range as obtained from the high-resolution run (Paper~I). There are some significant differences in the spatial distributions of these parameters of the ISM and ejecta components. For example, $n_{\rm e}t_{\rm ISM}$ is clearly centrally peaked, while $n_{\rm e}t_{\rm ejecta}$ has smoother distribution except for a ``dark belt'' in the SE half of the SNR interior and the much lower values at the NW rim which is known to be dominated by the shocked high density ISM (e.g., \citealt{Nikolic13}). The difference in $n_{\rm e}$ of the ISM and ejecta is even more significant, with the overall shape of the ISM component consistent with a limb-brightened shell-like structure, while the ejecta appears to be centerally peaked with a ``dark belt'' at the same location as the $n_{\rm e}t_{\rm ejecta}$ map (Fig.~\ref{PaperIIfig:paraimg}g,h). 

After correcting for the density distribution, $n_{\rm e}t$ can be converted to $t_{\rm ion}$ with time dimension, which in principle traces the ionization history of the plasma. $t_{\rm ion,ISM}$ peaks in the center, indicating that the forward shock propagates outward, and the outer shell of the ISM is shocked just recently. The ionization history of the ejecta, on the other hand, is more complicated. The general trend of the spatial distribution of $t_{\rm ion,ejecta}$ is opposite to that of $t_{\rm ion, ISM}$, i.e., $t_{\rm ion,ejecta}$ decreases from outer to inner regions. This indicates that the reverse shock propagates inward if it is the major ionization mechanism of the ejecta. The most significant peak of $t_{\rm ion,ejecta}$ locates between the NW shell and the soft X-ray brighter SE half of the SNR. This region has low density, but high abundance of heavy elements from both ISM and ejecta (e.g., Mg; see \S\ref{PaperIIsubsec:DiscussionAbundance}). It probably represents regions between the forward and reverse shocks where a turbulent mixture of the shocked ISM and ejecta occurs (as predicted by numerical simulations, e.g., \citealt{Ferrand12}). The central region of the SNR has the lowest $t_{\rm ion,ejecta}$, indicating that the ejecta in the innermost region is newly shocked.

\begin{figure}
\begin{center}
\hspace{-0.2in} 
\epsfig{figure=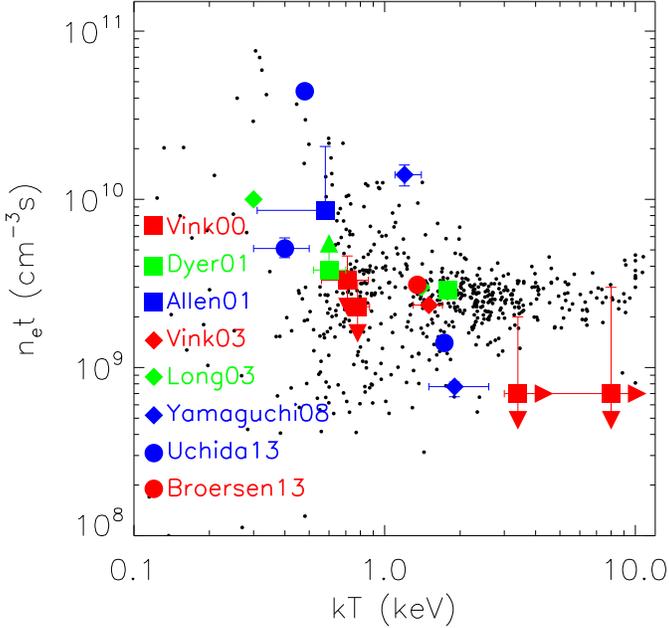,width=0.5\textwidth,angle=0, clip=}
\caption{$kT$ v.s. $n_{\rm e}t$ diagram. Black dots are the values of the ejecta component obtained from each region in the low-resolution run. Large colored symbols are obtained from the literatures as denoted in the lower left corner. For $kT\gtrsim2\rm~keV$, $n_{\rm e}t$ is in a narrow range, while for lower $kT$, $n_{\rm e}t$ is much more diverse.}\label{PaperIIfig:kTnet}
\end{center}
\end{figure}

We further compare the thermal and ionization states of the ejecta component to results from some previous works through the $n_{\rm e}t-kT$ diagram in Fig.~\ref{PaperIIfig:kTnet} \citep{Vink00,Vink03,Dyer01,Allen01,Long03,Yamaguchi08,Uchida13,Broersen13}. Compared to a similar $n_{\rm e}t-kT$ diagram constructed with the high-resolution run (Paper~I), the anti-correlation of $n_{\rm e}t$ and $kT$ caused by the degeneracy of these two parameters in spectral fitting is much less significant. This means the joint fitting of the higher signal-to-noise ratio spectra is much more reliable. We do not find too many regions with $kT<0.5\rm~keV$ as found in the high-resolution run (there is a peak at 0.22~keV on the PDFs of $kT$). This low temperature component is most likely produced by some high density knots of the ISM, so only present in some small regions which are not resolved in the low-resolution run. At high temperature ($kT\gtrsim2\rm~keV$), $n_{\rm e}t$ is quite stable in a narrow range of $(1.5-5)\times10^{9}\rm~cm^{-3}~s$. These data points mostly represent the shocked ejecta in the SE half of the SNR (Fig.~\ref{PaperIIfig:paraimg}a,g). On the other hand, many regions in the NW half of the SNR has a low temperature of $\sim1\rm~keV$ (Fig.~\ref{PaperIIfig:paraimg}a); the ionization states of them are quite diverse with $n_{\rm e}t$ in a range from $\sim5\times10^{8}\rm~cm^{-3}~s$ to $>10^{10}\rm~cm^{-3}~s$. The gradual decrease of $n_{\rm e,ejecta}$ from SE to NW (Fig.~\ref{PaperIIfig:paraimg}h) and the relatively large $t_{\rm ion,ejecta}$ (Fig.~\ref{PaperIIfig:paraimg}i) of these regions are consistent with the above scenario that they represent the turbulent mixing region between the shocked ISM and ejecta. Their relatively low temperature may be naturally explained by the adiabatic cooling caused by the fast expansion of this region.

\begin{figure*}
\begin{center}
\epsfig{figure=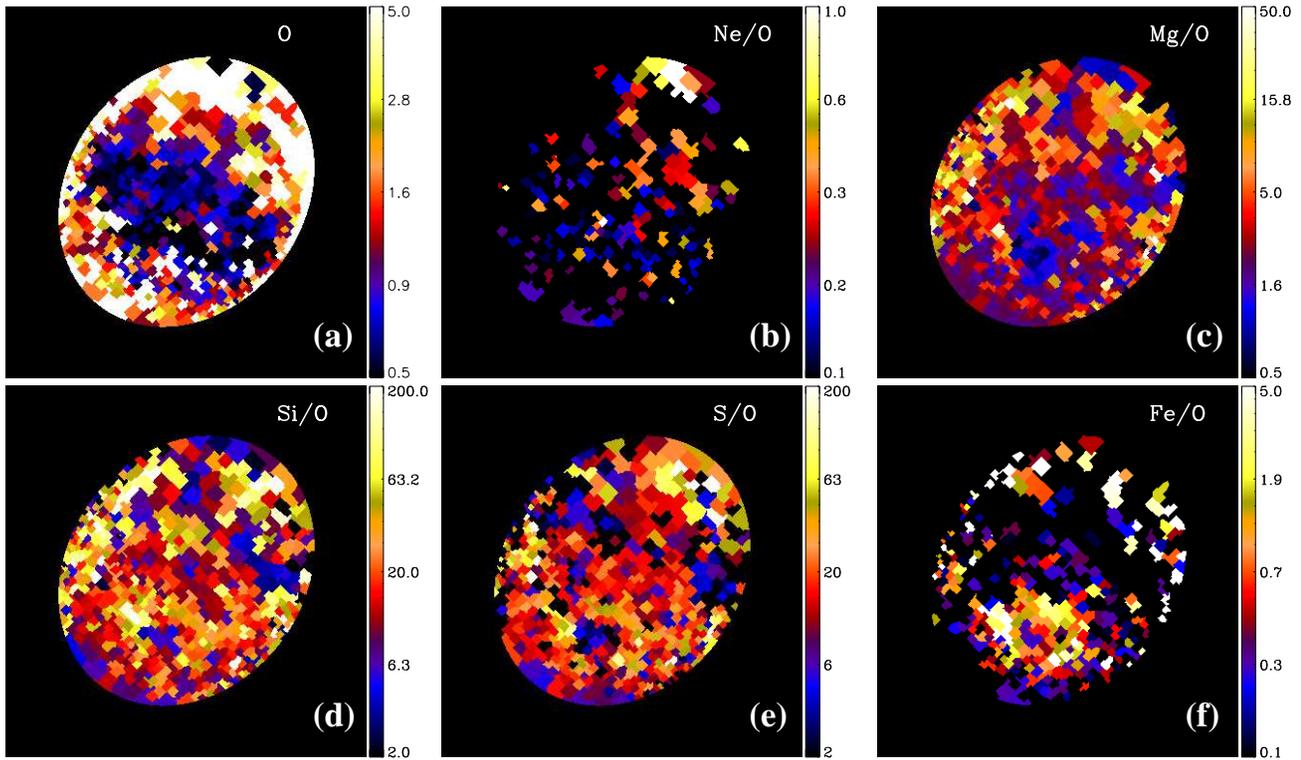,width=1.0\textwidth,angle=0, clip=}
\caption{Abundance maps of the ejecta component. The oxygen abundance (a) is the absolute abundance in unit of solar, while the abundances of other elements are the abundance ratios to oxygen, also in unit of solar values.}\label{PaperIIfig:abundanceimg}
\end{center}
\end{figure*}

The discussions presented in this section is mainly based on the qualitative trends shown in the parameter maps. To our knowledge, there is no dedicated 3D numerical simulations of SN1006 considering the particle acceleration and surrounding ISM distribution. Such hydrodynamical simulations, coupled with the NEI code (e.g., \citealt{Ferrand12,Orlando16}), could be quantitatively compared to the parameter maps shown in this paper. As a well-studied SNR in a relatively uniform environment, such a quantitative comparison between models and observations of SN1006 will play a key role in understanding the hydrodynamical evolution of Type~Ia SNRs.

\subsection{Metal abundances of the SN ejecta}\label{PaperIIsubsec:DiscussionAbundance}

The abundance or abundance ratio (to oxygen) maps of different elements are presented in Fig.~\ref{PaperIIfig:abundanceimg}. As discussed in \S\ref{PaperIIsubsec:SpecModel} and Paper~I, the absolute abundances of heavy elements are difficult to determine because of the degeneracy of abundance and EM in spectral fitting. Although this effect may significantly affect the measurement of the O abundance, the overall trend of O distribution of the ejecta component (Fig.~\ref{PaperIIfig:abundanceimg}a) is consistent with what we found in Paper~I through equivalent width (EW) and O abundane maps. The non-uniform and in many cases super-solar O abundance strongly indicate that O emission in SN1006 is not only from the ISM.

In contrast, the distribution of Ne in the ejecta is largely consistent with a primarily ISM origin (see Fig.~\ref{PaperIIfig:samplespec} for an example). In most of the regions, the Ne abundance of the ejecta is consistent with the lower limit set in spectral fitting. Even some strong Ne emitting features such as the NW shell and the ``dark belt'' revealed by the Ne EW and Ne abundance maps from the ``1-T'' model in Paper~I do not appear in the Ne abundance maps of the ejecta component here (Fig.~\ref{PaperIIfig:abundanceimg}b).

The Mg distribution is similar as O, but shows larger gradient, i.e., the Mg/O ratio shown in Fig.~\ref{PaperIIfig:abundanceimg}c is not uniform and similar as the O distribution. Therefore, similar as the O, there should also be some Mg contained in the ejecta. However, the O- or Mg-rich ejecta only appear in the NW half of the SNR representing the mixed ISM and ejecta between the forward and reverse shocks.

Si and S emission lines are mainly produced by the ejecta (e.g., Fig.~\ref{PaperIIfig:samplespec}). They also have similar spatial distributions (Fig.~\ref{PaperIIfig:abundanceimg}d,e). A significant difference between Si/S and O/Mg distributions is that the former is centrally filled, so probably have two components: one coincide with the O/Mg-rich ejecta (high abundance in the NW half) and the other one distributes in the inner region or SE half without much O/Mg.

\begin{figure}
\begin{center}
\hspace{-0.25in} 
\epsfig{figure=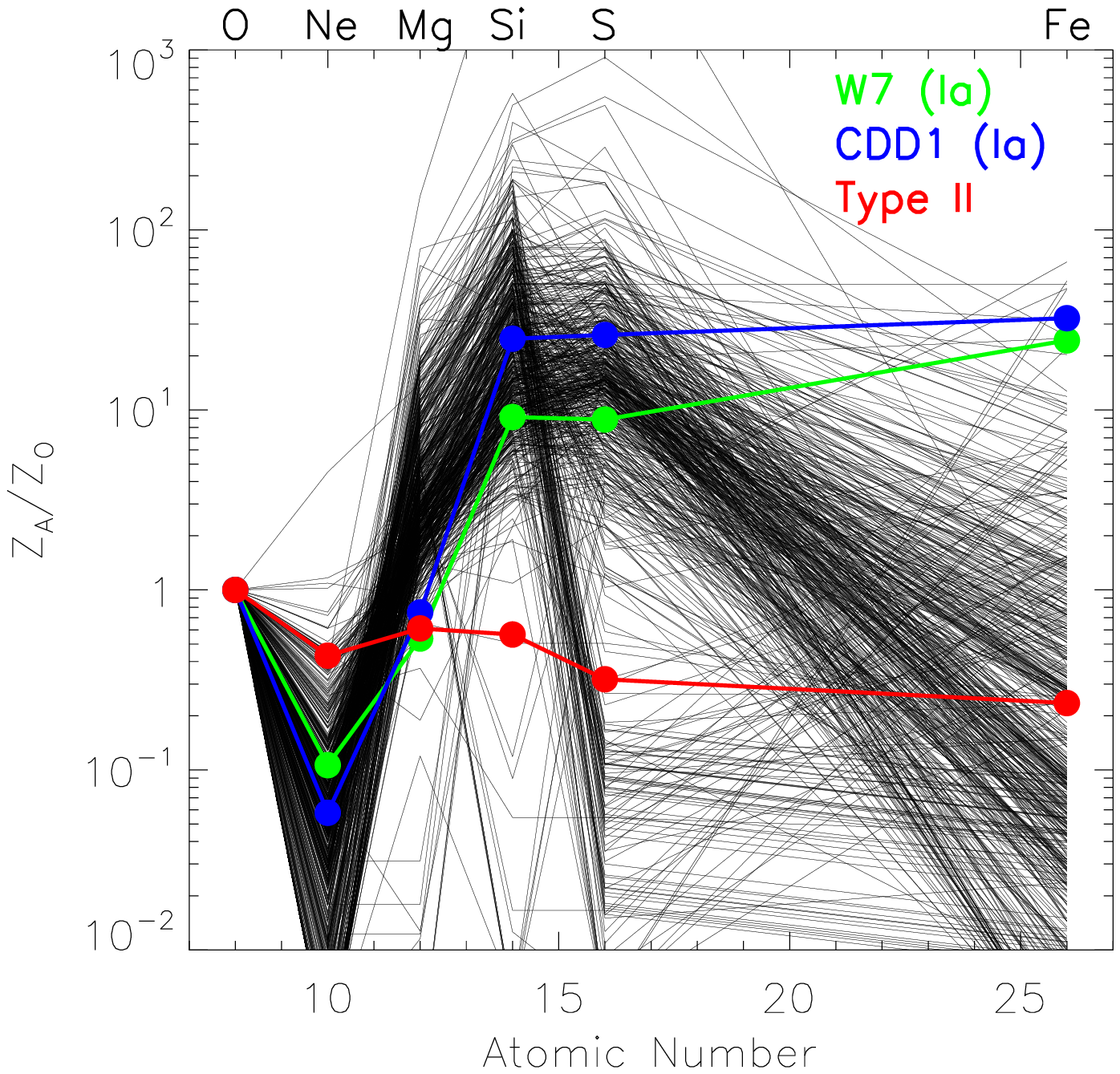,width=0.5\textwidth,angle=0, clip=}
\caption{The relative abundances (to O) of different elements (Ne, Mg, Si, S, Fe) of the ejecta component. Data from the same region are connected together with a thin solid line. Theoretical model predictions of Type~Ia (the classical deflagration model W7 and the delayed detonation model CDD1) and Type~II (10-50~$\rm M_\odot$) SNe from \citet{Iwamoto99} are plotted with large filled circles connected with thick solid lines. The O-Si abundances are in general consistent with the Type~Ia SN products, but the observed Fe and sometimes S abundances are significantly lower than predicted by standard Type~Ia SN yields.}\label{PaperIIfig:AbundancePattern}
\end{center}
\end{figure}

In most regions, the Fe abundance is consistent with the lower limit set in spectral fitting. Such a low Fe abundance may be biased as the major tracer of Fe emission is the L-shell lines around $\sim1\rm~keV$, which are weak at low ionization states. As a result, in the absence of Fe~K-shell lines in most of the regions, we may miss some Fe at low ionization states in the X-ray spectra (\ion{Fe}{2} absorption lines are found toward some UV-bright background sources; e.g., \citealt{Wu93}). In addition to the low ionization age of some shocked Fe, an alternative explanation of the low Fe abundance is that the Fe-rich ejecta may distribute in the inner region of the SNR and even not yet being reached by the reverse shock. This is indicated by the nearly shell-like Fe-rich feature with clearly super-solar Fe/O ratio (Fig.~\ref{PaperIIfig:abundanceimg}f). This feature also appears on the Si/O and S/O maps (Fig.~\ref{PaperIIfig:abundanceimg}d,e) and is likely a recently shocked feature with $t_{\rm ion,ejecta}$ probably slightly higher than the surrounding regions (Fig.~\ref{PaperIIfig:paraimg}i). The existence of both shocked and unshocked ejecta is further supported by UV absorption line studies of background sources (e.g., \citealt{Hamilton97}).

The spatial distributions of ejecta, as traced by the emission or absorption lines of some heavy elements in both X-ray and UV/optical, are clearly asymmetric (e.g., \citealt{Hamilton97,Winkler05,Yamaguchi08,Uchida13}; also see Paper~I and the above discussions). This asymmetric ejecta distribution could either be a result of asymmetric ambient ISM distribution (which could often produce strongly asymmetric outer shells and sometimes also asymmetric inner shells, e.g., \citealt{Chen08}), or be produced by an asymmetric explosion of the progenitor star. The ambient gas distribution, as traced by \ion{H}{1} 21-cm observation of the neutral gas \citep{Dubner02}, optical emission line observations of the ionized gas (e.g., \citealt{Winkler03,Nikolic13}), or IR observations of the interstellar dust \citep{Winkler13}, is clearly asymmetric. This asymmetric ISM distribution could explain the enhanced O, Mg, Si, and S emissions in the NW half of the SNR, which may be produced by the mixture of the shocked ISM and ejecta. The strong forward shock in this direction and possibly also the strong reverse shock may enhance the turbulent mixing of the post-shock gases. However, such an asymmetric ISM distribution cannot explain the strong Si, S, and Fe emission in the SE half of the SNR, where the preshock ISM density is much lower and the ejecta is most likely reverse shocked instead of mixing with the shocked ISM. Furthermore, the X-ray observation of the electron density distribution does not favor a strong asymmetry in post-shock density in the SE side (e.g., \citealt{Miceli09,Miceli12}, Paper~I). Therefore, the asymmetric SN explosion, instead of only asymmetric ISM distribution, is likely the major reason of the asymmetric ejecta distribution, as also suggested by the geometric models presented in other works (e.g., \citealt{Hamilton97,Winkler05,Uchida13}). 

In Fig.~\ref{PaperIIfig:AbundancePattern}, we compare the abundances of some heavy elements (Ne, Mg, Si, S, Fe) of the ejecta to the nucleosynthesis yields predicted by some theoretical models from \citet{Iwamoto99}. Heavy elements in SN1006 can be divided into two clearly distinguished groups: O-Ne-Mg and Si-S. These two groups are synthesized in different processes in nuclear burning: O-Ne-Mg from C burning while Si-S from O burning. Compared to core collapsed SNR, in Type~Ia SNR, the C burning products have much lower contribution to the abundance pattern in the ejecta than the O burning ones. Therefore, we conclude that the observed abundance patterns of the ejecta of SN1006 are consistent with standard Type~Ia SN models for O, Ne, Mg, Si, and S, while inconsistent with the Type~II SN model. However, the abundance of Fe, which is a product of Si burning, is always the lower limit in spectral fitting (as also shown in Fig.~\ref{PaperIIfig:abundanceimg}f) and inconsistent with the Type~Ia SN model. This also occurs in some regions for S as indicated by the low S abundance branch in Fig.~\ref{PaperIIfig:AbundancePattern}. The low abundance of Fe and S is most likely a result that a significant fraction of the SN ejecta, which is rich in Fe and sometimes S, has either a too low ionization state or not yet been reached by the reverse shock (also suggested in the analysis of UV absorption lines, e.g., \citealt{Hamilton97}).

\section{Summary}\label{PaperIIsec:Summary}

Based on the \emph{XMM-Newton} large program on SN1006 and our newly developed tools for spatially resolved spectroscopy analysis as described in Paper~I, we study the interior regions of SN1006 dominated by the thermal emission after excluding the two non-thermal limbs. We construct 583 tessellated regions (the low resolution run) with a higher signal-to-noise ratio (compared to the high resolution run presented in Paper~I) in order to resolve all the prominent soft X-ray emission lines, including the Si lines at $\sim1.8\rm~keV$. For each region, we jointly fit all the spectra extracted from different instruments and different observations with a ``2-T'' model plus the \ion{O}{7} K$\delta-\zeta$ lines, the non-thermal emission, and the background components. The two thermal components represent the ISM and ejecta contributions. The joint fitting with the ``2-T'' model significantly improves the spectral analysis, resulting in much lower $\chi^2/\rm d.o.f.$ values as compared to the ``1-T'' fitting of the stacked spectra in Paper~I. We then construct maps of various parameters and discuss their scientific implications. Our key results are summarized below.

The spatial distributions of the thermal and ionization states (traced by $n_{\rm e}t$, $n_{\rm e}$, and $t_{\rm ion}$) of the ISM and ejecta show some significantly different features. In general, these features are consistent with a scenario that the ISM (ejecta) is heated and ionized by the forward (reverse) shock propagating outward (inward). The low surface brightness region between the NW shell and the X-ray brighter SE half of the SNR most likely represents the region between the forward and reverse shocks where the shocked ISM and ejecta turbulently mix with each other. For regions with $kT_{\rm ejecta}\gtrsim2\rm~keV$, $n_{\rm e}t_{\rm ejecta}$ is in a narrow range; these regions represent the reverse shocked ejecta mostly in the SE half of the SNR. On the other hand, for regions with lower $kT_{\rm ejecta}$, $n_{\rm e}t_{\rm ejecta}$ is much more diverse, and may represent the mixed ISM and ejecta between the forward and reverse shocks.

Emission lines of different elements in and around SN1006 have different spatial distributions so probably different origins. Ne is mostly originated from the ISM, while a significant fraction of O and Mg are originated from the ejecta. Si and S are mostly from the ejecta. Most of the Fe-rich ejecta has not yet or just recently been reached by the reverse shock. This scenario is consistent with the above scenario for the ionization history. The abundance pattern of different elements in the ejecta is consistent with typical Type~Ia SN products. The overall spatial distribution of heavy elements, such as the enhanced ejecta emission in the SE half of the SNR, supports a scenario of asymmetric explosion of the progenitor star of SN1006. In addition, the asymmetric ISM distribution also plays an important role in shaping the soft X-ray emission, such as the enhanced O, Mg, Si, and S emissions from the ejecta between the NW shell and the soft X-ray brighter SE half of the SNR.

\bigskip
\noindent\textbf{\uppercase{acknowledgements}}
\smallskip\\
\noindent JTL acknowledges the financial support from NASA through the grants NNX13AE87G, NNH14ZDA001N, and NNX15AM93G.

\end{document}